\documentstyle[12pt]{article}
\begin{document} 
\begin{flushright}
Freiburg THEP-97/26\\
quant-ph/9711012
\end{flushright}
\begin{center}
\vspace*{1.0cm}

{\bf WIGNER FUNCTION AND DECOHERENCE}\footnote{To appear in
the Proceedings of the Fifth Wigner Symposium,
edited by P. Kasperkovitz
(World Scientific, Singapore, 1998).}

\vskip 1cm

{\bf Claus Kiefer}

\vskip 0.3 cm 

Fakult\"at f\"ur Physik \\ Universit\"at Freiburg \\
Hermann-Herder-Stra\ss e 3 \\ D-79104 Freiburg, Germany

\end{center}

\vspace{0.5 cm}

\begin{abstract}
I briefly review the role of the
Wigner function in the study of the quantum-to-classical transition
through interaction with the environment (decoherence).
\end{abstract}

\vspace{0.5 cm} 

My contribution to the Fifth Wigner Symposium in Vienna touches
upon two of Wigner's fields of interest:
His work on the interpretation of quantum theory \cite{wi1}
and his discussion of quasi-probability functions 
as encoded in the famous Wigner function \cite{wi2}.
As far as the interpretation is concerned, a crucial role
in \cite{wi1} is attributed to a result by Zeh \cite{zeh}
showing that it is usually very unrealistic to consider
systems as being isolated from their natural environment.
Except for small objects such as atoms and small molecules,
they are quantum-entangled with a huge number of other degrees
of freedom. This fact {\em must} therefore be taken into account
to gain a proper interpretation of quantum theory.

Under ordinary circumstances, these correlations with the
environment lead to the local appearance of classical properties --
a process known as decoherence (see the comprehensive
review \cite{dec}, where all necessary details can be found).
Amongst the most important features of decoherence is its
ubiquitous, unavoidable, occurrence as well as its
irreversible nature. Technically, decoherence is investigated
through the local, reduced, density matrix after the degrees of freedom
of the environment are traced out. It is important
to emphasise that the interaction with the environment
leads to the emergence of a distinguished local ``pointer basis"
which is stable in time and which {\em defines} the local property
that is called ``classical". The process of decoherence
has been experimentally monitored in quantum optics
experiments \cite{ha}.

While a very weak coupling to the environment is usually
sufficient to ensure classical behaviour, the local classical
equations of motion only hold approximately if this
coupling is not too strong. Otherwise, a large back reaction
may result, which can disturb the local dynamics or even
freeze it (Zeno effect). It is the discussion of this aspect
for which the Wigner function is a useful tool (see 
Sect.~3.2.3 of \cite{dec}).
It is used, in particular, to investigate the {\em correlations}
which exist between position and momentum after the interaction
with the environment is taken into account.

Some of the relevant examples are the following:
\begin{itemize}
\item {\em Localisation of objects.} This is an important example
because it shows why macroscopic objects 
appear localised within quantum theory.
 The interaction with the environment can prevent
the spreading of the wave packet even in cases where the mass is so
small that it would otherwise be noticeable (such as for a small
dust grain). 
\item {\em Decoherence of fields.} Macroscopic field strengths
can become classical through interaction with charged fields.
On the other hand, charges become classical quantities through
interaction with electromagnetic fields.
\item {\em Emergence of classical spacetime.} A quantised 
gravitational field would exhibit genuine quantum features.
Due to its universal coupling with all other fields in Nature,
the entanglement becomes so strong that at least the global
gravitational degrees of freedom (such as the radius of the
Universe) become classical to a high degree of accuracy.
The notion of an (approximate) classical time parameter
emerges through symmetry breaking \cite{dec}.
\item {\em Emergence of classical fluctuations in the
early Universe.} This is of utmost importance for the structure
formation. Quantum fluctuations arise in a natural way in the
context of inlationary universe models, but only if they become
classical can they be interpreted as providing the seeds for
galaxy formation \cite{PS,KP}.

\end{itemize}

In the above examples, one often encounters a wave function of the form
\[ \psi(x,y) \propto \exp\left(\mbox{i}S(x)\right)
    \exp\left(-\frac{\Omega(x)}{2}y^2
    \right)\ , \]
where $\Omega$ is complex, and $x$ ($y$) is an abbreviation for the
degrees of freedom of the system (environment). For example,
in the case of QED, $y$ may be the charged field degrees of
freedom, while $x$ may be the electric field. In many of the
applications, $\psi$ represents a squeezed state.

Integrating out the ``irrelevant" degrees of freedom $y$, one gets
the reduced density matrix, $\rho$, for the ``relevant" degrees $x$.
{}From $\rho$ one can then construct the corresponding Wigner function
in the standard way \cite{wi2,dec}. In the Gaussian approximation, the width
of $\rho$ gives a quantitative measure for the remaining coherence,
while the corresponding width in the Wigner function is a measure
for the correlation between position and momenta. Since one width
is the inverse of the other, an ``uncertainty relation" between
coherence and correlation holds.

For the above state, the correlation between position and momentum
is of the form $p_x=dS/dx +\ldots$, where $\ldots$ denotes the
back reaction of the $y$-degrees of freedom on the system $x$.
In the QED case one finds, for example, that a constant electric
field will be first slightly increased, but becomes weaker
after particle creation comes into play. Analogous results
hold for the above cited examples of the gravitational field
and the fluctuations in the early Universe. 

The topic ``Wigner function and decoherence" is thus of great
relevance in many branches of physics, covering
such different fields like particle physics,
quantum optics, and cosmology.

\end{document}